      [1999/12/01 v1.4c Il Nuovo Cimento]
\documentclass{cimento}

\def\kms{km~s$^{-1}$}
\def\etal{{\it et al.}}
\def\sqd{{deg$^{2}$}}
\def\arcmin{$^{\prime}$}
\def\arcsec{$^{\prime\prime}$}
\def\msun{$M_\odot$}

\usepackage{graphicx}  
\title{ALFALFA: an Exploration of the $\MakeLowercase{z}=0$ HI  Universe}
\author{Riccardo Giovanelli}
\instlist{\inst Dept. of Astronomy, Cornell University, Ithaca, NY 14853}
\PACSes{
\PACSit{95.35.+d,}{95.80.+p, }{95.85.Bh, }{98.52.Nr, }{98.52.Sw, }
{98.52.Wz, }{98.58.Ge, }{98.58.Nk, }{98.62.Py, }{98.62.Ve, }{98.80.Es}
}
\begin{document}

\maketitle

\begin{abstract}
The Arecibo Legacy Fast ALFA (ALFALFA) Survey is a program aimed at obtaining a 
census of HI-bearing objects over a cosmologically significant volume of the local 
Universe. It will cover 7074 square degrees of the high latitude sky accessible 
with the Arecibo 305m telescope, using the 7-beam feed L-band feed array (ALFA).
Started in February 2005, as of Summer of 2007 survey observations are 44\%
complete. ALFALFA offers an improvement of about one order of magnitude in
sensitivity, 4 times the angular resolution, 3 times the spectral resolution,
and 1.6 times the total bandwidth of HIPASS. Although it will cover only one
quarter the sky solid angle surveyed by HIPASS, ALFALFA will detect approximately
six times as many sources, with a median depth of 110 Mpc. Preliminary results of 
ALFALFA are presented, with emphasis on those related with the Virgo cluster.
\end{abstract}

\section{Introduction}
Baryons make up about 4.5\% of the mass/energy budget of the Universe, and only
1/6 of its matter density. At $z=0$ the vast majority of baryons are thought to
exist in the form of coronal and intergalactic gas, 
at temperatures $>10^{5}$ K; $\Omega_{stars}$ is a tiny 0.0027 and $\Omega_{cold~gas}$
an even smaller 0.0008, of which a bit over half is neutral Hydrogen~\cite{ref:omega}. 
This unimpressive budgetary datum could well prompt the question: why do we care
about HI? Several reasons for caring are relevant to the purview of this conference.
First, HI is easy to detect at 21 cm wavelength, most of the emission originates in
optically thin regions and cold gas masses are reliably measured; the abundance of
cold gas is a reliable indicator of star forming potential for an extragalactic
system. Second, the distribution of HI, which extends farther out than other easily
detectable components in a galaxy, makes it an excellent tracer of the large--scale
dynamics of its host. Third, scaling relations of disks, such as that between luminosity
and rotational width, make HI measurements good cosmological tools: for example in
the measurement of $H_\circ$, peculiar velocities, the convergence depth of the Universe
and the local matter density field. Fourth, because of its distribution at relatively
large galactocentric distances, HI is vulnerable to external influences and thus
constitutes a good tracer of tidal interactions, mergers and other environmental
effects. Fifth, it can be the dominant baryonic component in low mass galaxies and
thus provide a reliable census of low mass systems in the galactic hierarchy.
Hence, ALFALFA.

\section{What is ALFALFA?}

Wide angle surveys of the extragalactic HI sky became possible with the advent of 
multifeed front--end systems at L--band. The first such system with spectroscopic 
capability was installed on the 64~m Parkes telescope in Australia, and has produced 
the excellent results of the HIPASS survey \cite{ref:bar}. The 1990s upgrade 
of the Arecibo telescope optics made it possible for that telescope to host feed 
arrays, as proposed by \cite{ref:kil}. Eventually a 7-beam radio ``camera'',  
named ALFA (Arecibo L--band Feed Array), became operational enabling large--scale 
mapping projects with the great sensitivity of the 305--m telescope. 

ALFALFA will 
map the extragalactic HI emission at $cz<18000$ \kms ~over 7074 \sqd. Exploiting the large collecting 
area of the Arecibo antenna and its relatively small beam size ($\sim 3.5$\arcmin), 
ALFALFA will be eight times more sensitive than HIPASS with $\sim$four times 
better angular resolution. The combination of sensitivity and angular resolution 
allows dramatically improved ability in determining the position of HI sources, 
a detail of paramount importance in the identification of source counterparts at 
other wavelengths. Furthermore, its spectral backend provides 3 times better 
spectral resolution (5.3 \kms ~at $z = 0$) and over 1.4 times more bandwidth. These
advantages offer new opportunities to explore the extragalactic HI sky. A comparison 
of ALFALFA and other past and current HI surveys is given in Table 1. Data taking 
for ALFALFA was initiated in February 2005 and, in the practical context of time 
allocation at a widely used, multidisciplinary national facility, completion of the 
full survey is projected to require a total of 6 years.

\begin{table}[!ht]
\caption{Comparison of Blind HI Surveys}
\smallskip
\begin{center}
\begin{tabular}{cccccccc}
\hline
\noalign{\smallskip}
Survey & Beam      & Area   & res    & rms$^a$ & V$_{med}$ & N$_{det}$ & Ref\\
       & (\arcmin) & (\sqd) & (\kms) &   & (\kms)    &           &    \\
\noalign{\smallskip}
\hline
\noalign{\smallskip}
AHISS   & 3.3 &    13 & 16 & 0.7 & 4800 &   65 &  $^b$ \\
ADBS    & 3.3 &   430 & 34 & 3.3 & 3300 &  265 &  $^c$ \\
WSRT    & 49. &  1800 & 17 & 18  & 4000 &  155 &  $^d$ \\
HIPASS  & 15. & 30000 & 18 & 13  & 2800 & 5000 &  $^{e,f}$ \\
HI-ZOA  & 15. &  1840 & 18 & 13  & 2800 &  110 &  $^g$ \\
HIDEEP  & 15. &    32 & 18 & 3.2 & 5000 &  129 &  $^h$\\
HIJASS  & 12. &  1115 & 18 & 13  & $^i$ &  222 &  $^i$\\
J-Virgo & 12. &    32 & 18 &  4  & 1900 &   31 &  $^j$\\
AGES    & 3.5 &   200 & 11 & 0.7 & 12000&      &  $^k$\\
ALFALFA & 3.5 &  7074 & 11 & 1.7 & 7800 &$>$25000& $^l$\\
\noalign{\smallskip}
\hline
\end{tabular}
{\small

$^a$ mJy per beam uniformly referred at 18 \kms ~resolution;
$^b$ Zwaan \etal ~(1997, \textit{ApJ} 490, 173);
$^c$ Rosenberg \& Schneider (2000, \textit{ApJSS} 130, 177);
$^d$ Braun \etal ~(2003, \textit{AAp} 406, 829);
$^e$ Meyer \etal ~(2004, \textit{MNRAS} 350, 1195);
$^f$ Wong \etal ~(2006, \textit{MNRAS} 371, 1855);
$^g$ Henning \etal ~(2000, \textit{AJ} 119, 2696);
$^h$ Minchin \etal ~(2003, \textit{MNRAS} 346, 787);
$^i$ Lang \etal ~(2003, \textit{MNRAS} 342, 738), HIJASS has a gap in velocity coverage between 4500-7500
\kms, caused by RFI;
$^j$ Davies \etal ~(2004, \textit{MNRAS} 349, 922);
$^k$ Minchin \etal ~(2007, \textit{IAU 233}, 227);
$^l$ Giovanelli \etal ~(2007, \textit{AJ} 133, 2569).
}
\end{center}
\end{table}

Some of the main science goals to be addressed by ALFALFA which are of special 
relevance to these proceedings include: (i) the determination and environmental 
variance of the HI mass function, especially at its faint end and its impact on 
the abundance of low mass halos; (ii) the large--scale structure characteristics 
of HI sources, their impact on the ``void problem'' and metallicity issues; (iii)
providing a blind survey for HI tidal remnants and ``cold accretion''; (iv) determining
a direct characterization of the HI diameter function; (v) the ALFALFA survey area 
includes $\sim 2000$ continuum sources with fluxes sufficiently large to make useful 
measurements of HI optical depth, and hence to provide a low z link with DLA absorbers.

The minimum integration time per beam in seconds 
$t_s$ necessary for ALFA to detect an HI source of HI mass $M_{HI}$, width 
$W_{kms}$, at a distance $D_{Mpc}$ is
\begin{equation}
t_s \simeq 0.25 \Bigl({M_{HI}\over 10^6 
M_\odot}\Bigr)^{-2} (D_{Mpc})^4 
\Bigl({W_{kms}\over 100}\Bigr)^\gamma, \label{eq:ts}
\end{equation}
where the exponent $\gamma\simeq 1$ for $W_{kms}<300$ \kms and increases to 
$\gamma\simeq 2$ for sources of larger width. Thus the depth
of the survey, i.e. the maximum distance at which a given HI mass can be
detected, increases only as $t_s^{1/4}$.
A corollary of this scaling law is the fact that, once $M_{HI}$ is
detectable at an astrophysically satisfactory distance, it is more 
advantageous to maximize the survey solid angle than to increase the
depth of the survey through longer dwell times. In passing, we note that
the $t_s$ required to detect a given $M_{HI}$ at a given distance decreases
as the 4th power of the telescope diameter: Arecibo offers a tremendous
advantage because of its huge primary reflector. The
effective integration time per beam area of ALFALFA
is of order of 40 sec, which yields a minimum detectable HI mass
of $2\times 10^7$ $M_\odot$ at the distance of the Virgo cluster and
in a catalog of sources with a median $cz\sim 7800$ \kms. As of mid--2007, 44\% 
of the survey solid angle has been fully mapped. Further details on the design 
and progress of the survey can be seen in \cite{ref:gio1} and at the 
URL {\it http://egg.astro.cornell.edu/alfalfa}.
ALFALFA is an open collaboration. Anybody with a legitimate scientific
interest and willing to participate in the development of the survey can
join. Access to cataloged survey products
can be obtained at {\it http://arecibo.tc.cornell.edu/hiarchive/alfalfa}
and survey progress, guidelines for joining and other details can be obtained
at {\it http://egg.astro.cornell.edu/alfalfa}. 

\begin{figure}[ht]
\centerline{
\scalebox{0.7}{%
\includegraphics{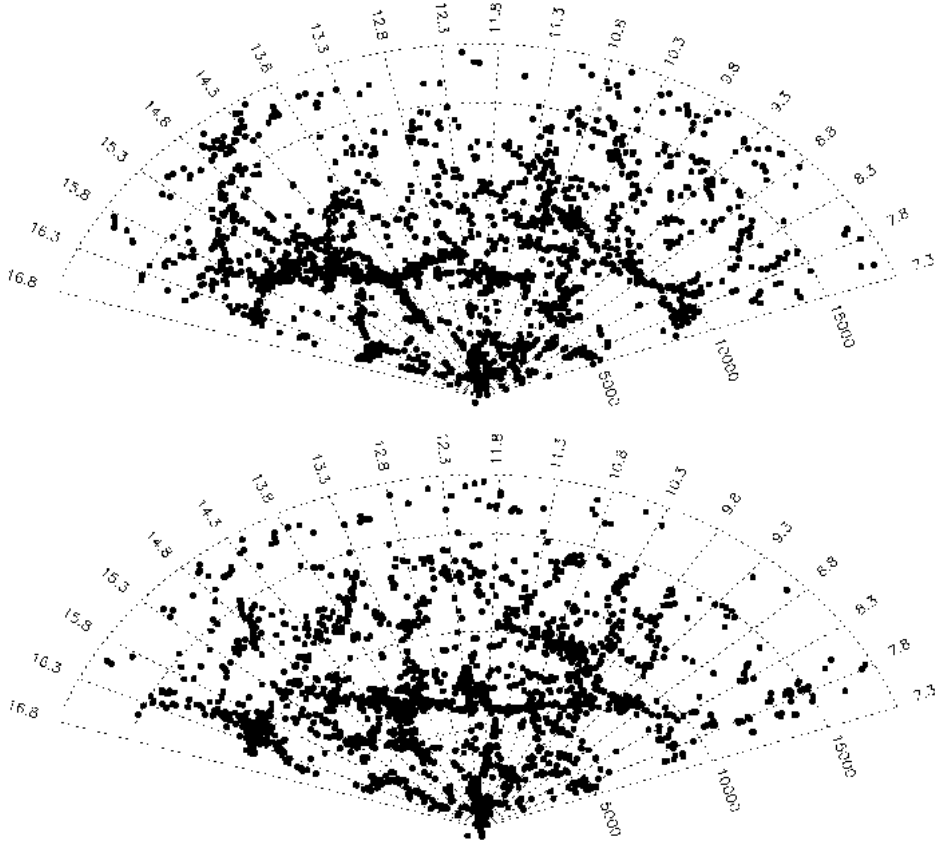}
}}
\vskip 1cm
\caption[]{\small
{Wedge plots of HI sources detected by ALFALFA. Both figures cover the same range in 
R.A.=[7.5$^h$--16.5$^h$]; the upper diagram corresponds to 2657 sources in the region
Dec=[12$^\circ$--16$^\circ$], whilee the lower one refers to 2580 sources in the 
contiguous region Dec=[8$^\circ$--12$^\circ$]. Together, they represents 15\% of the 
ALFALFA survey. Note that due to RFI, ALFALFA is effectively blind in the redshift range 
between approximately 15000 and 16000 \kms. 
}}\label{cones}
\end{figure}

Two catalogs of HI sources extracted from 3-D spectral data cubes have been 
accepted for publication in 2007 \cite{ref:gio2} 
\cite{ref:sai}, and several others are in preparation. As of Summer
of 2007, more than 5000 HI sources have been identified, over a solid angle
representing 15\% of the total ALFALFA survey. Figure \ref{cones} shows those
sources in the framework of two wedge plots, corresponding to contiguous declination
strips. While the two strips are non-overlapping, repeatability of the large
scale features bears witness to the correlation scale in the galaxy distribution.
Note that, due to the impact of local RFI, ALFALFA is
effectively blind in the redshift range $cz\sim$15000 to 16000 \kms.

\begin{figure}
\centerline{
\scalebox{0.13}{
\includegraphics{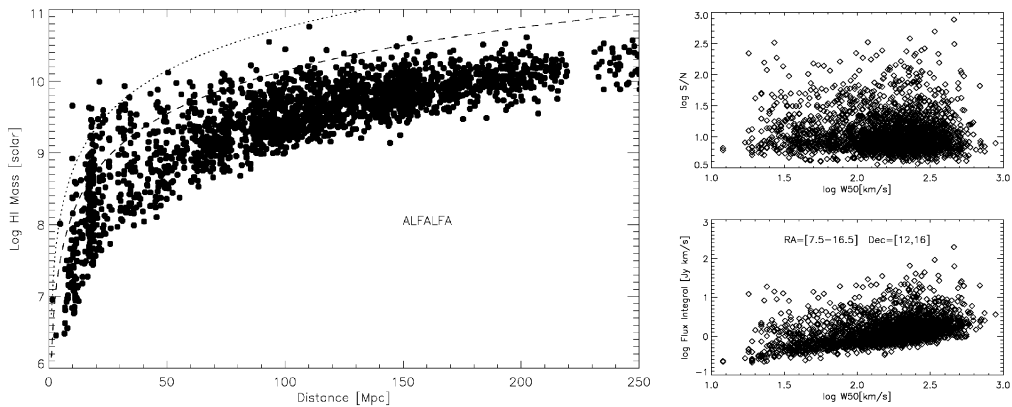}     
}}
\caption{
{\bf Left:} Sp\" anhauer plot HI sources in the region R.A.=[7.5$^h$--16.5$^h$],
Dec=[12$^\circ$--16$^\circ$]. The two smooth lines identify the completeness limit
(dotted) and the detection limit (dashed) for sources of 200 \kms ~linewidth for
the HIPASS survey. Note that due to rfi, ALFALFA is
effectively blind in the redshift range between approximately 15000 and 16000 \kms. 
{\bf Right:} Signal--to--noise ratio vs. velocity width (top) and Flux integral vs.
velocity width for the galaxies in the left--hand panel.
}\label{himd}
\end{figure}

Figure \ref{himd} shows a HI mass vs. distance diagram of the HI sources in
Figure \ref{cones}. Two smooth lines are overplotted, identifying respectively 
the completeness limit (dotted) and the detection limit (dashed) for sources 
of 200 \kms linewidth for the HIPASS survey. This diagram dramatically
illustrates the improvement ALFALFA represents, over previous surveys.
The median redshift of the catalog is $\sim$7800 \kms ~and its
distribution reflects the known local large scale structure. 
See Haynes' presentation in these proceedings for a discussion of the impact
of these observations on the faint end of the HI mass function.

For the same set of sources, the top panel on the right of Fig. \ref{himd} displays 
S/N vs. velocity width, while the bottom panel displays the flux integral
vs. velocity width. The quality of the ALFALFA signal extraction is apparent: 
the S/N of detections exhibits no significant bias with respect to velocity width.
Spectroscopic HI surveys are not single flux limited. The flux limit 
rises as $W^{1/2}$ for low velocity widths, changing to a linear rise for the
wider line profiles. Such a transition is observed near $\log W\simeq 2.5$. The
ALFALFA flux limit is $\sim 0.25$ Jy \kms ~for narrow lines, rising near 1 Jy \kms
~for the broadest ones.

The positional accuracy of HI sources is a very important survey parameter,
especially in the identification of HI sources with sources at
other wavelengths. The quality of 
the positional centroiding of a source depends roughly linearly on source S/N
and inversely on the telescope beam angular size. 
Consider, for example, a source barely detected by HIPASS at S/N$\simeq 6.5$.
The error box of its positioning will have a radius of approximately 2.5'.
The same source can be detected by ALFALFA with $S/N\simeq 50$; as the Arecibo
beam is about 4 times smaller than that of Parkes, the positional error box
for the ALFALFA observation is $\sim 0.1'$, thus making an optical identification
far more reliable. Positional accuracy of ALFALFA sources 
averages 24\arcsec ~(20\arcsec ~median) for all sources with S/N $>$ 6.5 and is 
$\sim$17\arcsec ~(14\arcsec ~median) for signals S/N $>$ 12. 

While in preliminary form, interesting statistical properties are starting to
emerge from the ALFALFA data, some of which are illustrated in M. Haynes's presentation
in these proceedings. The remainder of this paper is devoted to discussion
of HI sources with no obvious optical counterparts in the Virgo cluster region.

\section{VirgoHI21: a Dark Galaxy?}

What should we refer to as a ``dark galaxy'' (perhaps a misnomer)? A dark galaxy 
would be a starless halo, yet detectable at other than optical wavelengths, possibly 
in HI or through lensing experiments. Such objects are likely to exist, but hard to 
find. Within the CDM galaxy formation paradigm, such objects would have 
relatively low mass, were unable to form stars before re--ionization and
either lost their baryons or were prevented from cooling them thereafter,
by the IG ionizing flux. Yet we know of low mass galaxies in the Local Group which
not only made stars early on, presumably before re--ionization, but they 
were also capable of retaining cold gas and make stars at later cosmic times. 
Why then should we not expect the existence of low mass systems that were unable
to form stars but have retained baryons and have been able to cool them,
as the IG ionizing flux rarefies? We have extremely little observational
evidence for the existence of such systems. The SW component of the system 
known as HI1225+01 \cite{ref:che}
has $M_{HI}/L_{opt}>200$ and exhibits evidence for dynamical independence 
(a very small amplitude rotation curve) from the NE component, which has an optical counterpart. 
However, the SW component is not an isolated object and it cannot be excluded 
that it originated from a high speed tidal encounter of the NE component with 
a now remote passer--by, as the system lies in the outskirts of the Virgo cluster.
In that case the velocity gradients interpreted as a rotation curve may just be
tidal. The burden on observers is that of finding isolated systems resembling HI1225+01SW.

\begin{figure}
\centerline{
\scalebox{0.35}{
\includegraphics{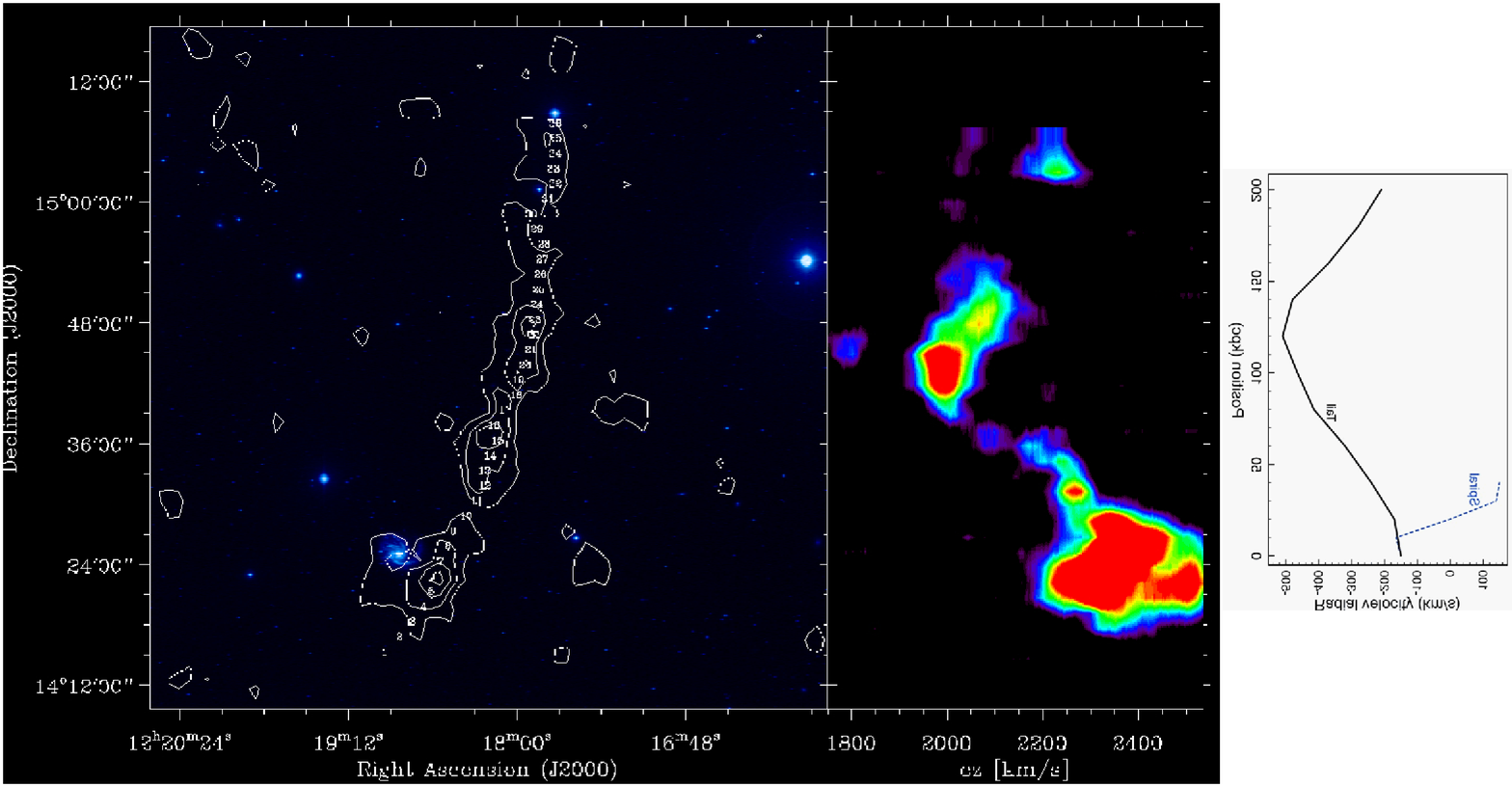}    
}}
\caption{{\it Left}: HI column density contours extracted from the ALFALFA
survey dataset, superposed on the SDSS image and centered on the position
of Virgo~HI21 (Minchin \etal ~2005a). {\it Center}: The velocity
of the HI emission peak as seen along the ridge of the stream. Note that
HI emission from NGC 4254 is excluded from the map to the left, but it 
is included in the one on the center image. {\it Right}: Position-velocity
line along the stream as modelled by Duc \& Burneaud, flipped and scaled
to match the two color images. VirgoHI21 was identified as a section of the 
HI stream extending from $14^\circ 41'$ to $14^\circ 49'$.}
\label{virgohi21_pv}
\end{figure}

VirgoHI21 was discovered at Jodrell Bank, corroborated by Arecibo and WSRT 
observations (\cite{ref:min} and refs therein). It lies some 100 kpc N of 
NGC4254, in the NW periphery of the Virgo cluster, projected $\sim 1$ Mpc from 
the cluster center and has a relative velocity of more than 1000 \kms with
respect to the cluster. Because of its large separation from optical galaxies
and the gradient seen in its velocity field, it was interpreted by its
discoverers as a dark galaxy. The ALFALFA data suggest a different 
scenario. The left--hand panel in Figure \ref{virgohi21_pv} displays
ALFALFA contours of HI flux, superimposed on an optical image, showing a gas
streamer extending some 250 kpc N of NGC 4254. The velocity field of the 
stream, which matches the velocity of NGC4254 to the S, is shown on the
center panel of the figure. The observations by the previous group did not reveal the
HI stream in its full extent: what they called VirgoHI21 is the bright section 
of the HI stream extending from  $14^\circ 41'$ to $14^\circ 49'$. The HI
mass in the disk of NGC2454 is $4.3\times 10^9$ M$_\odot$ and that associated with 
the stream is $5.0\pm0.6\times 10^8$ M$_\odot$. One of the driving arguments
for the interpretation of VirgoHI21 as an isolated disk
galaxy is the gradient seen in the velocity field\cite{ref:min}; ALFALFA data shows
that gradient to be just a part of the varying, large--scale velocity field
along the stream.

NGC 4254 is a system well known for its prominent $m=1$ southern spiral
arm. It is reasonable to postulate that this special feature is related
with the existence of the stream. Note the following:
\begin{itemize}
\item NGC 4254 moves at a large velocity with respect  to the cluster
($>1000$ \kms) and lies at a projected distance of $\sim$ 1 Mpc from M87.
\item The prominent m=1 arm is visible in the gas and in the disk stellar
population: gravity, rather than hydro phenomena such as ram pressure, is
at work.
\item The HI mass in the stream is only $\sim 10$\% of that in the NGC 4254
disk: the disturbance of NGC 4254 is relatively mild (it would not, in fact
be classified as an HI deficient galaxy).
\item The velocity field of the stream shows the coupling of the tidal
force and the rotation of NGC 4254, which suggests an interesting
timing argument:
\begin{enumerate}
\item the stream exhibits memory of a full rotational cycle of the NGC 4254 disk;
\item from the NGC 4254 VLA map of \cite{ref:pho}, we can get
the present outer radius of the HI disk (18.5 kpc) and the rotational
velocity at that radius (150 \kms); from those we compute a rotation
period of $\simeq 800$ Myr.
\end{enumerate}
\item Hence we estimate that the tidal encounter which gave rise to the
stream initiated some 800 Myr ago, a time comparable with the cluster
crossing time. If the interaction resulted from a high speed (of order
of 1000 \kms, the velocity differential between NGC 4254 and the cluster) 
close encounter with another galaxy and/or the cluster potential, the culprit 
for the tidal damage would now be $\sim 1$ Gpc away.
\end{itemize}

\begin{figure}
\centerline{
\scalebox{0.50}{
\includegraphics{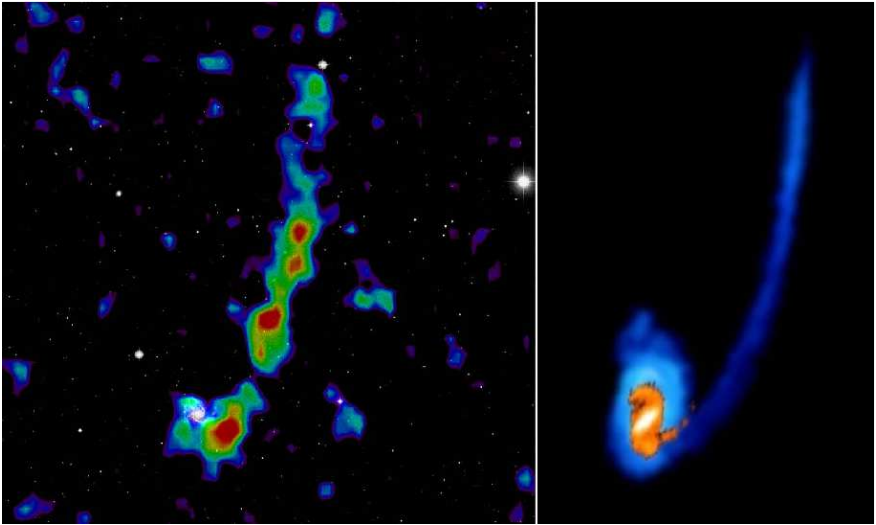}  
}}
\caption{{\it Left}: HI column density contours extracted from the ALFALFA
survey dataset, superposed on the SDSS image and centered on the position
of Virgo~HI21 \cite{ref:min}. {\it Right}: The stream as modelled 
by \cite{ref:duc}. VirgoHI21 was identified as a section of the HI stream 
extending from $14^\circ 41'$ to $14^\circ 49'$ (see preceding figure).}
\label{virgohi21}
\end{figure}

We conclude that the most reasonable interpretation of the system is that
of a relatively mild episode of harassment, resulting from the high
speed passage of NGC 4254 through the cluster periphery. These results are
discussed in greater detail in \cite{ref:hay}.

Duc \& Bournaud \cite{ref:duc} have produced a computer simulation of a high speed 
encounter of NGC 4254 with another peripheral cluster galaxy. The simulation 
matches extremely well both the morphology and the velocity field of the stream. 
A comparison of the model and the ALFALFA data is shown in figures \ref{virgohi21}
(stream morphology) and \ref{virgohi21_pv}, right hand panel (position--velocity).
The culprit responsible for the harassment of NGC 4254 could now be located
far from NGC 4254. The authors of the simulation speculate that, given its location 
and velocity, the culprit could be M 98 = NGC 4192. ALFALFA finds an extended
HI appendage apparently emanating from that galaxy.

The overall evidence for VirgoHI21 to be part of the phenomenology associated
with a tidal episode of harassment, rather than an isolated ``dark galaxy''
is thus quite strong.

\section{The NGC 4532/DDO 137 System}

\begin{figure}
\centerline{
\scalebox{0.36}{
\includegraphics{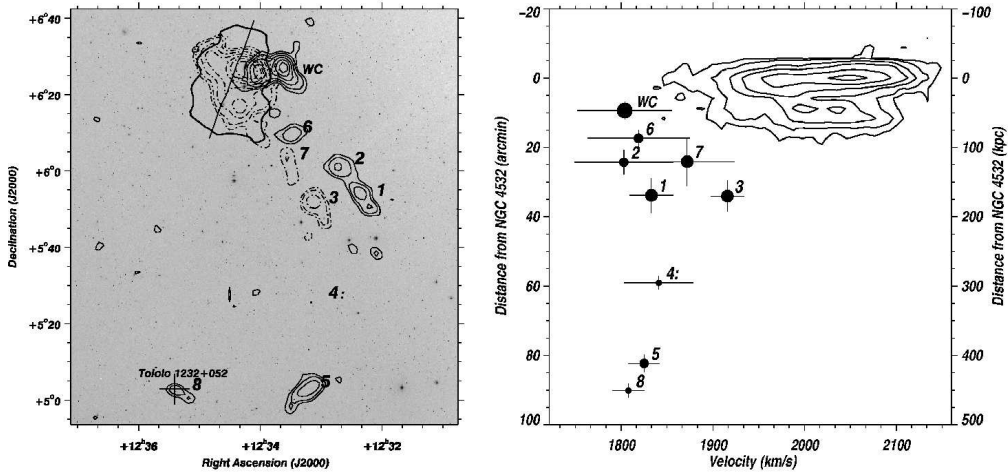}     
}}
\caption{{\it Left}: The bold contour at 0.3 Jy beam$^{-1}$ km s$^{-1}$, integrated 
over 1951 - 2139 km s$^{-1}$, encompasses the approximate area of the HI envelope
detected by Hoffman \etal (1993). Solid contours show tail emission integrated 
over 1784 - 1836 km s$^{-1}$, while dashed contours show emission integrated 
over 1868 and 1930 km$^{-1}$. {\it Right}: Position (radial distance from NGC 4532) 
as a function of velocity for the NGC 4532/DDO 137 system. The contours follow 
a locus (the solid line in the left panel) along the major axis of NGC 4532 and 
through the HI envelope of the pair system.}
\label{koop}
\end{figure}

This pair of galaxies is located to the South of the Virgo cluster. The two
galaxies are of late type (SmIII/SmIV) and very gas rich. Arecibo observations 
(\cite{ref:hof1} ~and refs. therein) revealed that some of the HI in the system is well
beyond the optical disks of the two galaxies. VLA observations\cite{ref:hof2}~
confirmed those results. ALFALFA maps expand on both of those results, revealing the presence
of cold gas at significantly larger galactocentric distances than previously realized.
The left panel of figure \ref{koop} covers a solid angle of about 3 \sqd, with a
bold solid contour approximately tracing the outer envelope of the HI gas reported
by \cite{ref:hof1}. Within that region, ALFALFA\cite{ref:koo}
detects an HI mass of 6.2 x 10$^9$
\msun, in agreement with previous reports. An additional 1.3 x 10$^8$ \msun~ is 
contained within a partially resolved clump $\sim$ 20$'$ west of NGC 4532, labelled
`western clump' (WC). A set of discrete clumps, numbered 1 through 8, outline the
ridges of two streams, connected by low column density emission not displayed
by the contour plots. They have HI masses varying between $2.5\times10^7$ and
$6.8\times10^7$ \msun. Clumps 1, 2, 4, 5 and 8 outline one  stream, while clumps
7 and 8 appear to be on a separate stream. The right-hand panel of figure \ref{koop}
provides an elementary position-velocity map of the system, showing the streamer
system apparently emerging from the low velocity side of NGC 4532 (solid contours).
No apparent optical counterparts are seen associated with WC and the numbered
clumps in the system; the overall HI mass associated with them is about 5 to $7\times10^8$
\msun, which is approximately 10\% of the galaxy pair's HI mass, a ratio similar to
that of the stream--to--galaxy in the NGC 4254 system discussed above. At the Virgo
cluster distance, the stream system associated with NGC 4532/DDO 137 extends 
over 500 kpc. As in the case of NGC 4254, the disturbance is spectacular, but
the ``damage'' caused to the galaxy pair appears to be mild. The size, velocity and 
other characteristics of the system suggest, again, a galaxy harassment scenario.
Modelling of the interaction that may have caused the streams is underway.

\section{A Cloud Complex near NGC 4424}

\begin{figure}
\centerline{
\scalebox{0.5}{
\includegraphics{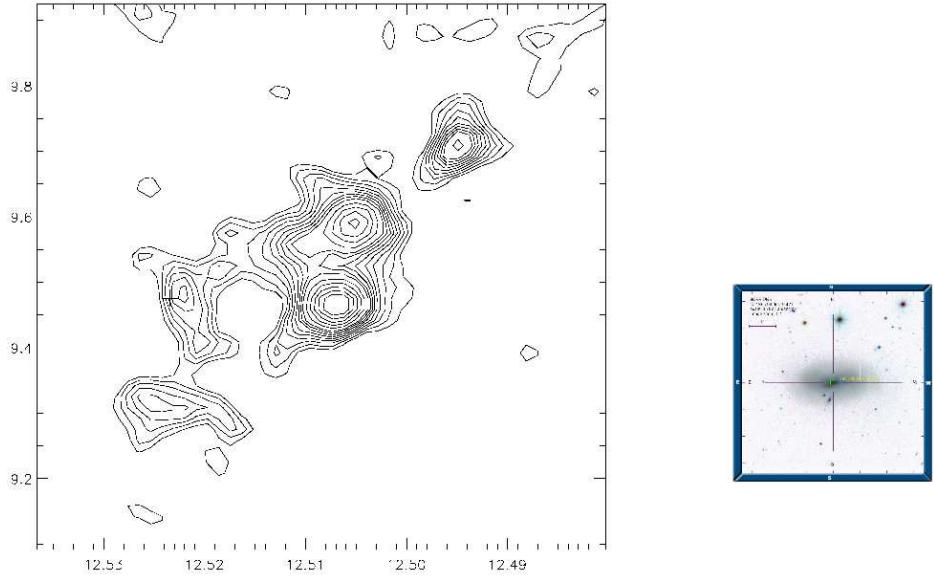}     
}}
\caption{Zeroth moment map of an HI cloud complex detected in the vicinity of NGC 4424.
The lowest contour plotted corresponds to approximately 0.8 mJy beam$^{-1}$. An SDSS 
image of NGC 4424 is shown at its approximate location with respect to the cloud
complex, although the optical image is enlarged by a factor of 2.5. The heliocentric
velocity of NGC 4424 is 441 \kms, while those of the clumps in the complex vary
between 485 and 609 \kms.}
\label{complex}
\end{figure}

Roughly halfway in the sky between M87 and M49, ALFALFA detects a conspicuous 
complex of HI clouds, shown in figure \ref{complex}. The nearest optical galaxy
with a velocity near that of the complex is NGC 4424, located some 40' W of the
complex center. An optical image of NGC 4424 is shown in figure \ref{complex},
at its approximate sky location with respect to the cloud complex. The optical 
image is however enlarged by a factor of 2.5 with respect to the features shown
in the HI map. The 
velocities of the individual clouds are (S to N) 476, 490, 601, 605 and 527 \kms, 
and their velocity widths are 48, 66, 45, 257 and 120 \kms. 
NGC 4424 has heliocentric velocity of 441 \kms. At the Virgo cluster distance, 
the individual clouds in the complex have HI masses between $0.4\times 10^7$ \msun 
~(to the SE) and $2\times 10^8$ \msun. The total HI mass of the complex is 
$\simeq 5\times 10^8$ \msun. The HI mass of NGC 4424 is $1.7\times 10^8$ 
\msun.  Given its HI mass and optical size, NGC 4424 is very HI deficient
($Def\simeq 1$, corresponding to having lost most of its cold gas\cite{ref:cor}).

Stretching over 200 kpc (at the Virgo cluster distance), the cloud complex does 
not appear to be gravitationally bound. With cloud--to--cloud velocity differences
of order of 100 \kms, the mean cloud separation will double over $\sim 1$ Gyr.
The complex thus appears to be a transient phenomenon. Plausible interpretations
of its nature are: (a) detached ISM from a single galaxy, either by ram pressure
or tidal forces; (b) group of mini halos falling in the cluster for the first time .
The absence of conspicuous (given the velocity widths involved) optical counterparts 
argues strongly (b). The HI deficiency and other properties of NGC 4424 are 
is strongly suggestive of environment--driven mechanisms at work, and likelihood
of association with the cloud complex. Occam's razor does not favor the idea of a 
cluster of dark galaxies, albeit the possibility that some of the clumps may give 
rise to the formation of tidal dwarfs is an attractive hypothesis.

\section{Other Virgo Features}

NGC 4192 = M 98 is a galaxy located in the NW periphery of
the Virgo cluster, with a negative $cz_\odot=-142$ \kms)
indicative of large relative motion with respect to the cluster itself.
A number of HI clouds with velocities between 60 and 100 \kms ~are
found in its vicinity, spectrally well separated from Milky Way
emission. They would typically be cataloged as High (or Intermediate)
Velocity Clouds, perigalactic (or Local Group) dwellers, and
most of them may very well be just that. Some of the clouds, however,
appear to stream out of M 98, stretching over more than $1^\circ$,
well matched both spatially and kinematically to the disk of M 98. 
The characteristics of this system are under close scrutiny. Of
particular interest is the  fact that \cite{ref:duc} suggest that M 98 may be the culprit
responsible for the harassment of NGC 4254, observed in the form of 
the stream VirgoHI21 is a part of.

\begin{figure}
\centerline{
\scalebox{0.5}{
\includegraphics{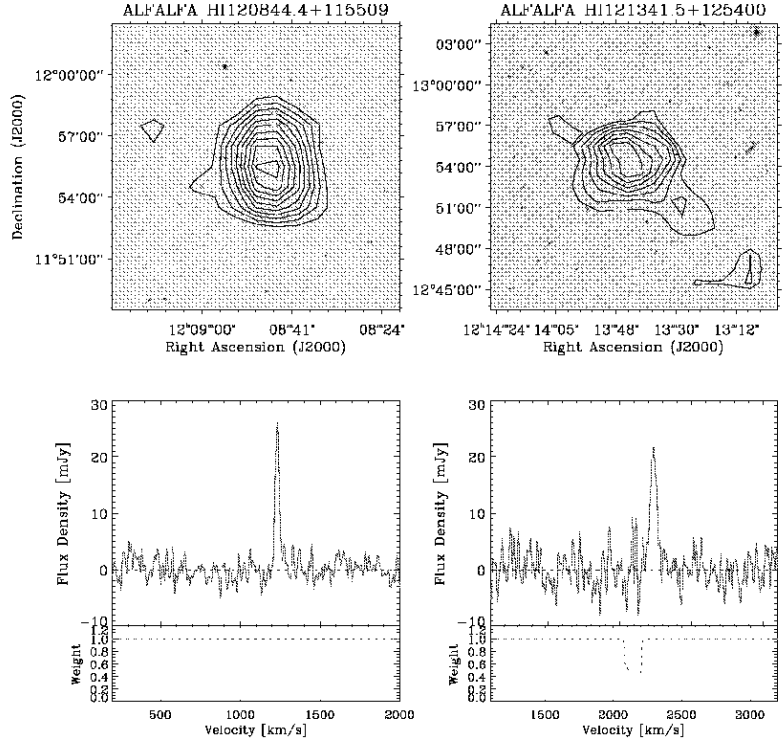} 
}}
\caption{HI spectra and contour maps of 2 sources in the Virgo cluster region
(see text for details).}
\label{cl2}
\end{figure}

ALFALFA has detected several other features lacking obvious counterparts,
as tabulated in \cite{ref:ken}. Figure \ref{cl2} shows two such examples:
the HI sources are unresolved by the Arecibo beam and thus extend fewer
than $\sim 15$ kpc at the Virgo distance, at which their HI masses would
be $4\times 10^7$ (left panel) and $8\times 10^7$ \msun (right panel); the 
second object may however be part of the M cloud, in the background of the 
cluster. No clear clues are available as to the nature of these objects.
They are sufficiently removed from the cluster center that ram pressure
effects are small. While the possibility that they may be primordial
remains open, the likelihood is high that they may constitute tidal remnants, 
the consequence of events such as those discussed in the previous sections.

\section{An Overall HI View of the Virgo Cluster}

\begin{figure}
\centerline{
\scalebox{0.5}{
\includegraphics{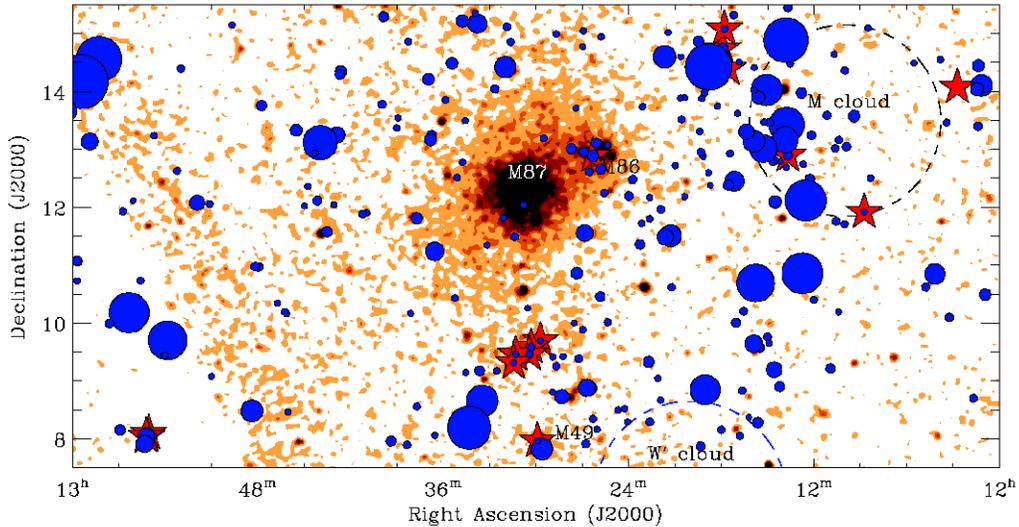}     
}}
\caption{Composite of the Virgo cluster: X ray emission (orange), HI sources
(blue circles) and HI sources with  no obvvious optical counterpart (red stars).
See text for details.}
\label{virgo}
\end{figure}

The HI content and extent of HI disks of galaxies that venture in the inner
regions of clusters have been known to be strongly affected(\cite{ref:cha},
\cite{ref:gio3},\cite{ref:hay2},\cite{ref:cay},\cite{ref:chu},\cite{ref:sol}).
Figure \ref{virgo} ~clearly illustrates the matter. About 200 HI sources are
detected by ALFALFA in the Virgo region shown in that graph; they are
identified by blue circles, the area of which is proportional to the HI mass,
varying between $2\times 10^7$ and $3\times 10^9$ \msun. The orange--to--red
contours represent the intensity of the X--ray emission, as imaged by ROSAT
\cite{ref:sno}. In HI, the Virgo cluster appears as a ring of emission
surrounding the X--ray emiting IGM. Sources detected in the inner parts
of the cluster typically correspond to highly HI deficient galaxies. The
red stars in the graph show the locations of HI sources found by ALFALFA
not to have obvious optical counterparts. The vast majority of those lie
in the outer parts of the cluster and are possibly remnants of tidal
events. With a complete census of the HI sources in the cluster --- a goal
nearly at hand --- and a fair understanding of the cluster structure and
dynamics, it will soon be possible to estimate the frequency and longevity of
environment driven events in the nearest cluster to us. It is interesting
to conclude with the emerging realization that the wealth of optically
inert sources found in the vicinity of the cluster does not appear to be
matched in other regions for which ALFALFA mapping is becoming complete.

{\bf }

\acknowledgments
This work has been supported by NSF grants AST--0307661,
AST--0435697, AST--0607007. The Arecibo 
Observatory is part of the National Astronomy
which is operated by Cornell University under
a cooperative agreement with the National Science Foundation.
Grazie to Guido Chincarini and Paolo Saracco for their kind invitation.

\end{document}